\documentclass[a4paper,fleqn]{cas-dc}

\usepackage[authoryear]{natbib}
\usepackage{cleveref}

\def\tsc#1{\csdef{#1}{\textsc{\lowercase{#1}}\xspace}}
\tsc{WGM}
\tsc{QE}
\tsc{EP}
\tsc{PMS}
\tsc{BEC}
\tsc{DE}

\begin{document}

\def\aj{AJ}
\def\apj{ApJ}
\def\apjl{ApJLett}
\def\apss{Astroph.Sp.Sci.}
\def\aap{A\&A}
\def\mnras{MNRAS}
\def\apjs{ApJS}
\def\apjsupp{ApJS}
\def\nat{Nature}
\def\memsai{Mem. SAI}
\def\iaucirc{IAU Circ.}
\def\apjl{ApJLett}
\def\aaps{A\&AS}
\def\aapr{A\&A Rev}

\let\WriteBookmarks\relax
\def\floatpagepagefraction{1}
\def\textpagefraction{.001}
\shorttitle{Properties of dwarf nova EY Cyg}
\shortauthors{Nabizadeh et~al.}

\title [mode = title]{X-Ray properties of dwarf nova EY Cyg and the companion star using an {\it XMM-Newton} observation}                      



\author[1]{Armin Nabizadeh}[type=editor,
                        orcid=0000-0002-2967-5402]
\cormark[1]
\ead{armin.nabizadeh@utu.fi}


\address[1]{Department of Physics and Astronomy, FI-20014 University of Turku, Finland}


\author[2]{S\"olen Balman}[%
   orcid=0000-0001-6135-1144]
\ead{solen.balman@gmail.com}

\address[2]{Currently Self-Employed, Nadide str. 26/2, Sisli, Istanbul 34381, Turkey}



\cortext[cor2]{Correponding Author}


\begin{abstract}
We present the X-ray analysis of dwarf nova EY Cyg using the 45 ks \textit{XMM-Newton} observatory archival data obtained in quiescence. We find orbital modulations in X-rays. We simultaneously fitted EPIC pn, MOS1 and MOS2 data using a model for interstellar medium absorption (\textit{tbabs}) and a multi-temperature plasma emission model with a power-law distribution of temperatures (CEVMKL) as expected from low accretion rate quiescent dwarf novae. The \textit{XMM-Newton} EPIC spectra of the source yields a maximum temperature $kT_{\rm max}$ $\sim$ 14.9$^{+3.3}_{-2.2}$ keV with an unabsorbed X-ray flux and luminosity of (1.8--2.0) $\times$ 10$^{-12}$ ergs$^{-1}$ cm$^{-1}$ and (8.7--9.7) $\times$ 10$^{31}$ ergs$^{-1}$, respectively, in the energy range 0.1 to 50 keV. There is 3--4 sigma excess at energies below 0.5 keV, we model the excess using \textit{MEKAL}, POWERLAW and BBODY models and favor the model \textit{MEKAL} which is physical. According to previous studies, the secondary in this system is thought to be a K-type star which may radiate in the soft X-ray region. The fit with an additive \textit{MEKAL} model gives a temperature of $kT$ $\sim$ 0.1 keV with an unabsorbed X-ray flux and luminosity of (2.7--8.8) $\times$ 10$^{-14}$ ergs$^{-1}$ cm$^{-1}$ and (1.3--4.2) $\times$ 10$^{30}$ ergs$^{-1}$, respectively, for the companion star. Based on the results from the timing and spectral analysis, we highly suggest that the secondary of EY Cyg is a K-type star.

\end{abstract}




\maketitle

\section{Introduction}

Cataclysmic variables (CVs) are interacting binary systems containing a white dwarf as their compact object which accretes material from a late-type low mass main sequence star \citep{warner2003cataclysmic}. Regarding the magnetic field strength of the white dwarf, CVs can be classified into magnetic and nonmagnetic systems. Magnetic CVs (MCVs) are those systems in which the magnetic field of the white dwarf is strong. Therefore, the white dwarf accretes matter through the magnetospheric lines \citep[][and references therein]{2012MmSAI..83..578M}. In nonmagnetic CVs, where the magnetic field of the white dwarf is weak enough (B < 0.01 MG), the accreting material forms an accretion disk around the white dwarf. In this work, a subclass of non-magnetic cataclysmic variables, dwarf nova (DN), has been considered. 

In dwarf nova systems, a continuous flow of matter is transferred to the Roche lobe of the white dwarf. The mass transfers at a low rate which is quiescence, and is interrupted every few weeks to tens of months by intense accretion flows. These enhanced accretion episodes, so-called outbursts, are of 2--9 mag amplitude and last from days to weeks. In general, the matter in the inner disk which has initially a Keplerian velocity is expected to dissipate its kinetic energy in order to accrete onto the slowly rotating white dwarf. This would create a boundary layer in the innermost region of the disk. The boundary layer is the connecting area between the accretion disk and central compact object with a radial extent. The standard accretion disk theory predicts that half of the accretion luminosity originates in the boundary layer in a range a few $\times$ 10$^{30}$--10$^{34}$ erg s$^{-1}$ and the other half emerges from the disk in the optical and ultraviolet (UV) wavelengths \citep[][]{1974MNRAS.168..603L, 1997A&A...327..602V,2012MmSAI..83..578M}. During the quiescent state (low-mass accretion rates), dwarf novae are mostly emitting in the hard X-rays \citep[][and references therein]{1985ApJ...292..535P,Balman2015}. A hot optically thin boundary layer is considered as the main source of these hard X-rays heated to temperatures of $\sim$10$^{8}$ K. The observations of dwarf novae in quiescence have readily shown this component, and only few have shown the expected soft component in the high states \citep{kuulkers2006compact,mukai2017x}. Note that if the rate of accretion is high, the boundary layer is expected to be optically thick which has a blackbody temperature of  $\sim$10$^{5}$ K radiating in soft X-rays or EUV \citep[][]{popham1995accretion}.  

More recently, \citet{Balman2012} have shown that the accretion flows in the inner disks of DN in quiescence are not optically thick and portray nonstandard hot flow characteristics. The power spectra with break frequencies in a range 1--6 mHz indicating the change in the flow to a nonstandard flow. In addition, the cross-correlation studies of simultaneous UV and X-ray light curves of DN yielding 90-180 sec time delays in the X-rays (X-rays lag UV emission) supports this outcome. \citet{Balman2019} reviews the characteristic flow structure in quiescent DN using the broadband noise characteristics in comparison with DN systems in outburst and also to magnetic CVs and other X-ray binaries. In addition, \citet{Balman2014} and \citet{balman2014swift} show that some nova-like systems and an old novae in quiescence that are at high accretion rates, do not show the soft X-ray component expected from the standard accretion theory, but shows hard X-rays with virialized flows. This also shows the existence of
nonstandard hot flow structure in the inner disks of other CVs in the X-ray emitting region. 


\begin{table}[width=\linewidth,cols=4,pos=t]
\caption{System Physical Parameters of EY Cyg.}
\label{tab:1}
\begin{tabular*}{\tblwidth}{@{} LLL@{} } 
\toprule
	
		Parameter & Unit & EY Cyg\\
		\hline
		
		\textit{Mass of WD} & \textit{M$_{\odot}$}&  1.10$\pm$0.09 \\
		\textit{Mass of secondary} & \textit{M$_{\odot}$} &  0.49$\pm$0.09  \\
		\textit{Inclination}& \textit{deg} &  14$\pm$1\\
		\textit{Period} & \textit{day} &  0.4593249(1) \\
		\textit{Ephemeris (T$_{0}$)} & \textit{HJD} & 2449255.3260(9)\\
		\textit{Distance} & \textit{pc} &  637$\pm$8\\
		\textit{Minimum V magnitude} &  &  14.8 \\
		\textit{Maximum V magnitude} &  &  11 \\

\bottomrule
\end{tabular*}
\end{table}

EY Cygni was first detected with ROSAT X-ray observatory in the energy range of 0.4--2.2 keV \citep{1992IAUC.5680....2O}. It is a U Gem type star with an orbital period of 11.0237976 h derived from absorption line analysis where the ephemeris was $T_{0}$ = 2449255.3260(9) \citep{2007A&A...462.1069E}. The physical parameters of the system are summarized in the Table~\ref{tab:1}. The visual magnitudes of the system during quiescence and outburst are $\sim$14.8 and $\sim$11, respectively \citep{2007A&A...462.1069E}. The analysis of long term AAVSO light curves revealed that the recurrence time of the outbursts in the system is about 2000 days \citep{2002AIPC..637...72T} which is 8 times longer than the previously measured cycle by \cite{1978JAVSO...6...60P}. Early studies of EY Cyg show that it contains a white dwarf with a mass of $M_{\rm wd}$ = 1.26 and a secondary of $M_{\rm sec}$ = 0.59 $M_{\odot}$. Later, \citet{2007A&A...462.1069E} using photometric analysis derived the masses as $M_{\rm wd}$ = 1.10 $\pm$ 0.09 $M_{\odot}$ and $M_{\rm sec}$ = 0.49 $\pm$ 0.09 $M_{\odot}$ confirming that the white dwarf in EY Cyg is massive. The results are also consistent with \citet{1998AAS...192.8206C} and \citet{2004AJ....128.1795S}. \citet{2007A&A...462.1069E} also found that the system has an inclination angle of 14 degrees. The spectral type of the secondary in this system is not fully determined yet. However, \citet{1962ApJ...135..408K} estimated that the spectral type of the secondary to be K0V. Later, \citet{1997MNRAS.287..271C} in a survey of 22 objects using spectral analysis with ISIS (Intermediate-dispersion Spectrograph and Imaging System) triple-beam spectrograph classified the secondary of EY Cyg to be in the range K5--M0. 

In far-UV spectral analysis of EY Cyg using FUSE and Hubble Space Telescope (HST) observations, \citet{2004AJ....128.1795S} shows that  EY Cyg contains white dwarf with $T_{eff}$ = 22000 K and an accretion belt with $T_{eff}$ = 36000 K. To do the analyiss, they estimated an average distance to the system as 450 pc. However, recently, the GAIA DR2 determined an accurate distance of 637 $\pm$ 8 pc for the system \citep{Harrison2018ApJ}.


\begin{figure}
	\centering
		\includegraphics[width=1.07\linewidth]{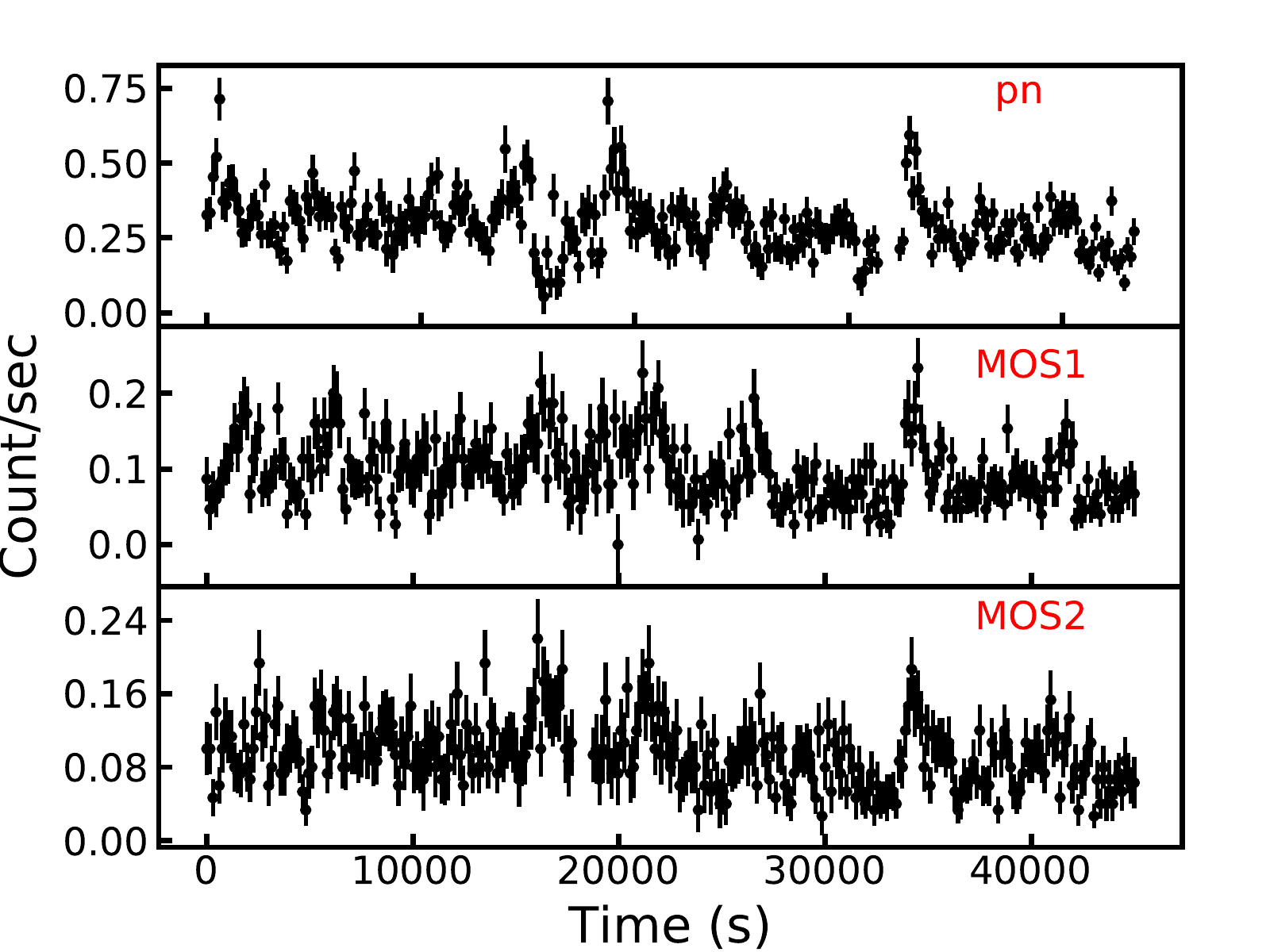}\vspace{0.5cm}
		\includegraphics[width=1.0\linewidth]{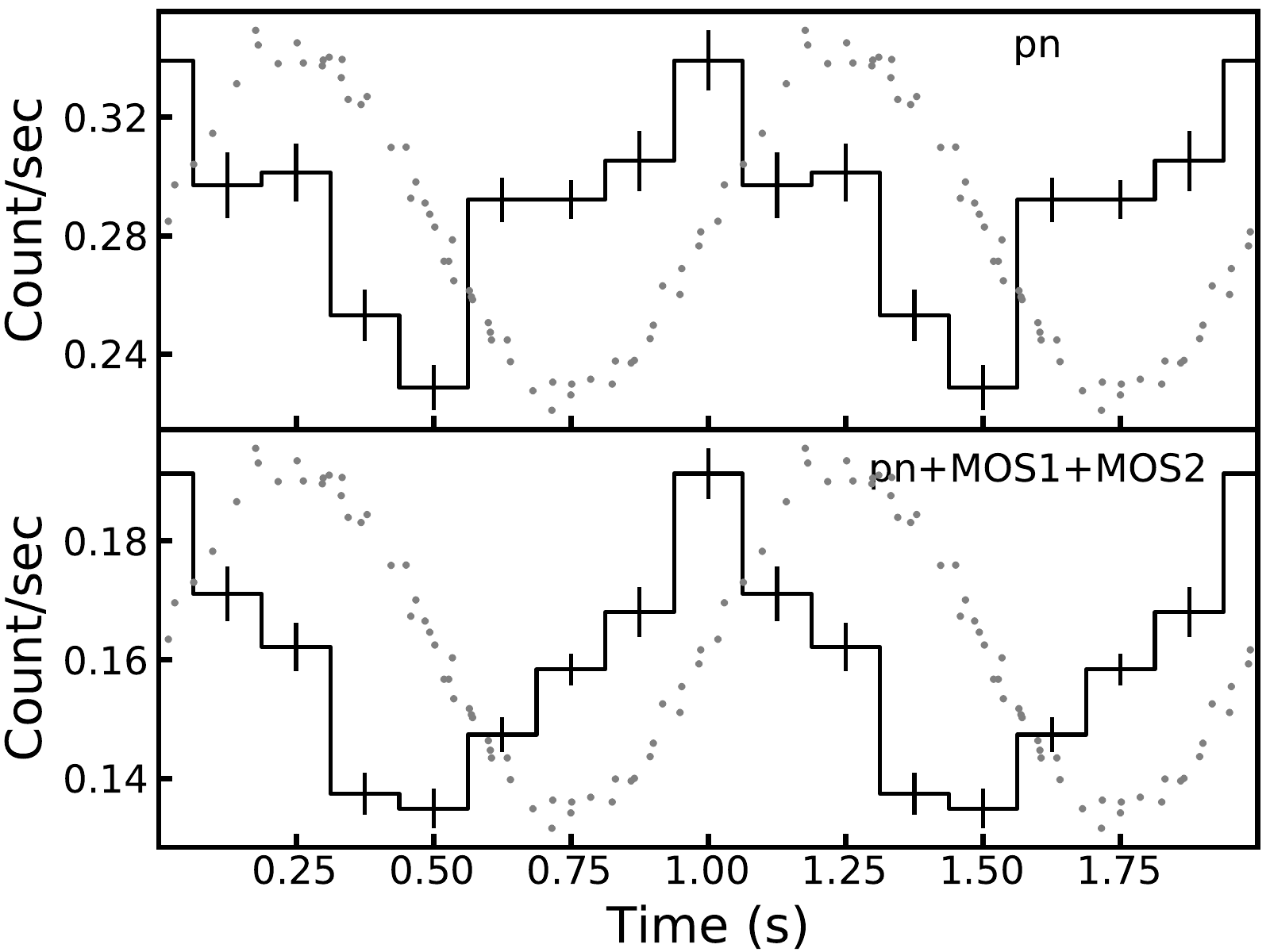}

	\caption{\textit{Top:} The \textit{XMM-Newton} EPIC pn, MOS1 and MOS2 light curves of EY Cyg plotted with the bin time of 150 s. \textit{Battom-top panel:} The EPIC pn light curve folded at the orbital period of 11.0237976 h.  \textit{Battom-lower panel:} The combined EPIC (pn+MOS1+MOS2) light curves folded at the orbital period of 11.0237976 h. The grey dots in each panel represent the radial velocities derived from absorption lines for 2005 June 26--July 1 adopted from Table 7 in \citet{2007A&A...462.1069E}.}
	\label{fig:1}
\end{figure}

The type of the secondary star in EY Cyg as well as the properties of the system in X-ray regime have not been studied until now. In  this work, we use the single {\it XMM-Newton} observation in the X-rays to determine the physical properties of the system and also to search for the type of the companion star.

\section{Observations and Data}

The X-ray Multi-Mirror Mission, \textit{XMM-Newton} satellite, \citep[][]{2001A&A...365L...1J} is a space X-ray observatory which was launched in December 1999. It carries three medium spectral resolution X-ray telescopes each with an European Photon Imaging Camera (EPIC) at the focus. There are also two Reflection Grating Spectrometers (RGS) for high resolution spectroscopy mounted behind two of the EPIC telescopes \citep{2001A&A...365L...7D}. In addition to them, there is also a 0.3 m optical/UV imaging telescope on-board. The optical monitor (OM) is a photon-counting instrument which enables simultaneous X-ray and optical/UV observations for following light curves and/or timing analysis or for imaging purposes \citep[][]{2001A&A...365L..36M}. EY Cyg was observed by \textit{XMM-Newton} observatory on 23 April 2007 (Observation ID 0400670101) with a 45 ks exposure time using the three X-ray instruments. Two multi-object spectrometer CCDs (EPIC MOS1 and MOS2) were operating in small-window mode \citep[time resolution $\sim$0.3 s;][]{2001A&A...365L..27T}, and EPIC pn was in the full-frame mode  \citep[time resolution $\sim$73 ms;][]{2001A&A...365L..18S}. There is no time series data for the source and RGS spectrum does not show features due to low S/N in the spectral bins. 

The archival \textit{XMM-Newton} data were reduced and analyzed using the \textit{XMM-Newton} Science Analysis System {\scriptsize SAS} version 11.0.0 with the latest available calibration files. In order to use well-calibrated and cleaned data, the standard filtering expressions were applied to all observations. For the pn data, single and double pixel events with {\scriptsize PATTERN}$\leqslant$4 and {\scriptsize \#XMMEA\_EP} and for the MOS, {\scriptsize PATTERN}$\leqslant$12 and {\scriptsize \#XMMEA\_EM} were used. The source spectra and the light curves were extracted from circular regions with radius of 18, 18.5  and 17.5  arcsec for pn, MOS1, and MOS2, respectively. The sizes were chosen to have the best signal to noise ratio for each detector. The background spectra and light curves, likewise, were extracted from source-free regions on the same chips. In addition, we used {\scriptsize FLAG = 0} to exclude bad pixels and events at CCD edges. The source observation was checked for effects of pile-up.


\section{Analysis and Result}

\subsection{Timing Analysis}
To carry out the timing analysis we first applied the solar system barycentric correction to the light curves. Then the standard {\scriptsize SAS} task {\scriptsize EVSELECT} was used to extract three light curves from the data set of all three X-ray CCDs with a bin size 0.1 s. All the light curves are background subtracted. Fig.~\ref{fig:1}-top shows the unfolded background-subtracted X-ray EPIC light curves (EPIC pn, MOS1 and MOS2) plotted with a bin time of 150 s. The light curves show EY Cyg in quiescence. To study the variation of the X-ray light curves, we combined all the background-subtracted light curves (EPIC pn, MOS1 and MOS2) and folded it at the orbital period of the system (11.0237976 h) using 8 phase bins (Fig.~\ref{fig:1}-bottom). The folding procedure were carried out with {\scriptsize XRONOS} task {\scriptsize EFOLD}. We used the ephemerides $T_{0}$ = 2449255.3260(9) + 0.4593249(1)$E$ HJD where the zero phase corresponds to the inferior conjunction of the secondary star \citep{2007A&A...462.1069E}. The EY Cyg folded mean light curve indicates a sinusoidal behavior of orbital modulation with a minima and maxima at phases 0.5 and 1.0, respectively. Since the accumulated phase error was very small (0.002), we were able to lock the phase of the X-rays to the optical light curves. Therefore, in order to make a comparison, we plotted the radial velocities derived from absorption lines for 2005 June 26--July 1 \citep[Table 7 in ][]{2007A&A...462.1069E} which shows a 0.25 phase offset with respect to our X-ray folded light curve. Since the absorption line radial velocity curves follow the secondary, where the lines are formed, and the X-rays follow the primary star and are produced close to the WD in the inner disk, such an offset is typical of comparisons between radial velocity and X-ray mean light curves.

\begin{figure}
	\centering
		\includegraphics[width=1.21\linewidth]{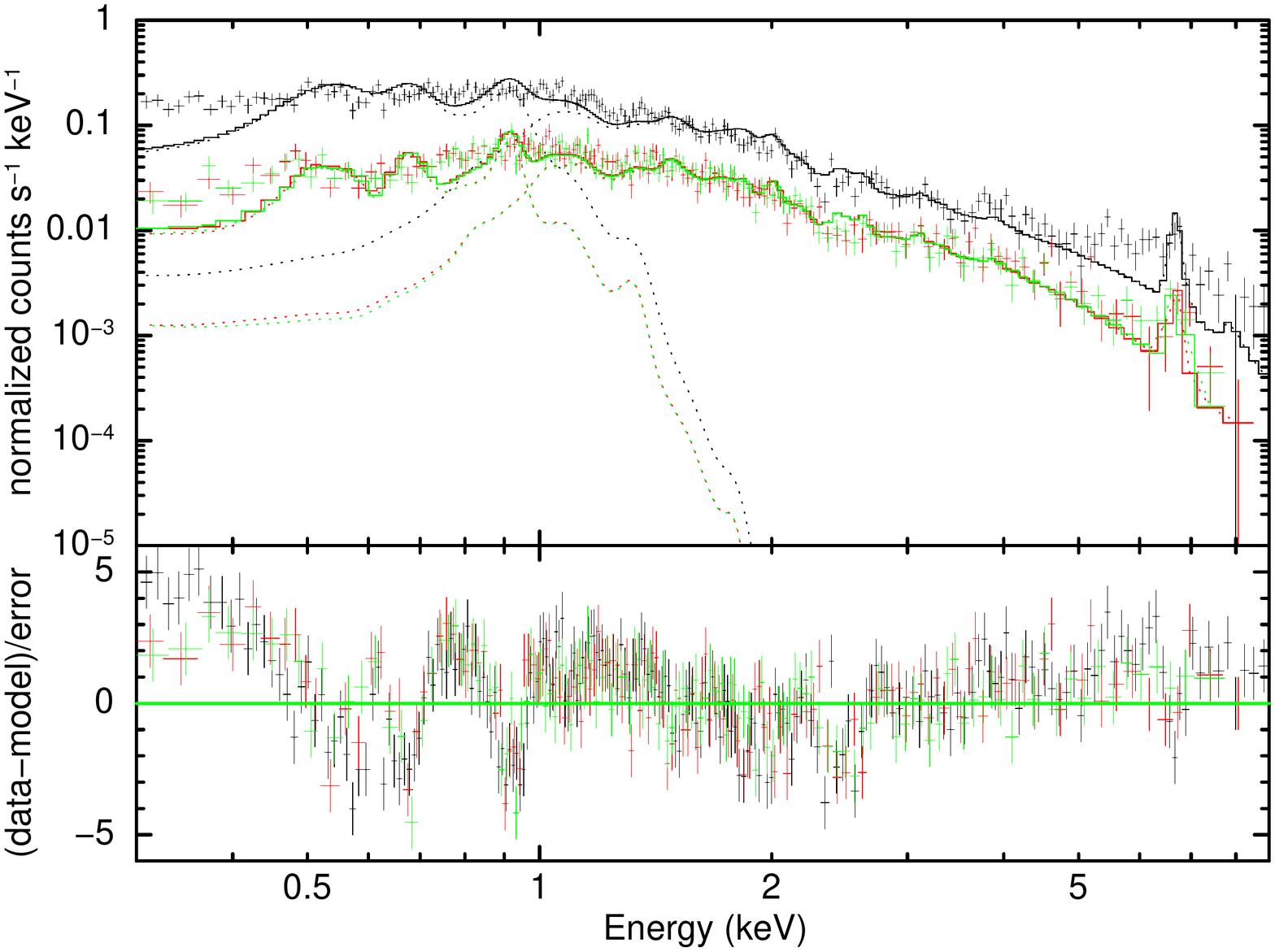}\vspace{-1.0cm}
		\includegraphics[width=1.15\linewidth]{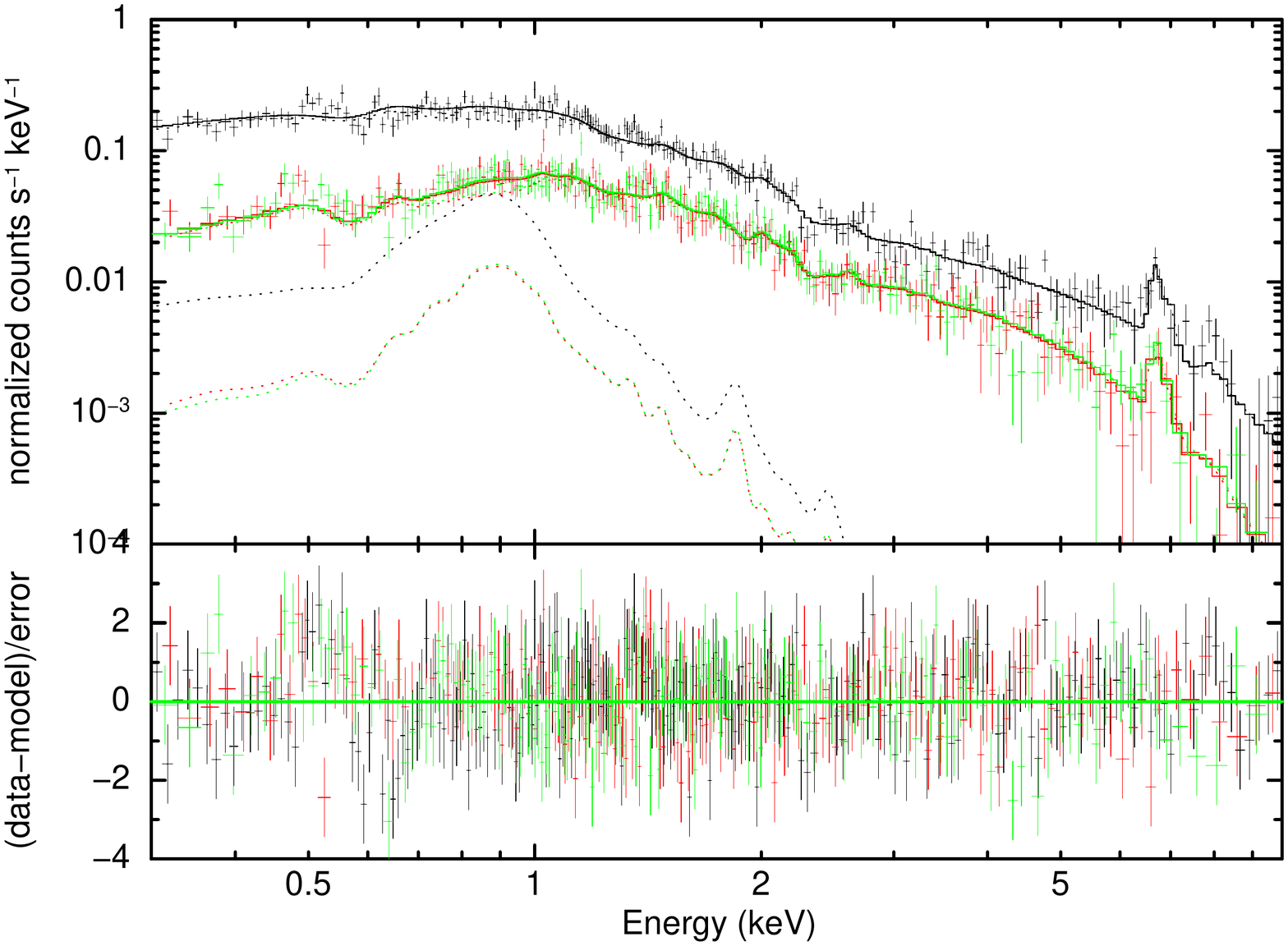}

	\caption{\textit{Top: }the combined EPIC pn, MOS 1 and MOS 2 spectra of EY Cyg together with the composite model [(tbabs$\times$constant$\times$(cevmkl)] fitted to the spectra. \textit{Bottom: }The combined EPIC pn, MOS 1 and MOS 2 spectra of EY Cyg together with the best-fit model [(tbabs$\times$constant$\times$(mekal+cevmkl)] fitted to the spectra. Lower panel in each plot shows the residuals in standard deviations.}
	\label{fig:2}
\end{figure}

\subsection{Spectral Analysis}

We reduced the event files and applied the {\scriptsize SAS} task {\scriptsize ESPECGET} in order to calculate the X-ray spectra for EPIC pn (0.35 count/s), EPIC MOS1 (0.1 count/s) and EPIC MOS2 (0.1 count/s) together with the appropriate response matrices and ancillary files. In order to increase the signal to noise ratio for good statistics in spectral bins, the spectra were grouped to have 55 counts in each energy bin for EPIC pn and 30 counts for MOS1 and MOS2. The spectral fitting was simultaneously performed for all three EPIC spectra using {\scriptsize XSPEC} version 12.8.0 \citep{1996ASPC..101...17A}. The photons with the energies below 0.3 keV and higher than 9.0 keV were ignored. We fitted the spectra with proper plasma emission models which are explained in detail below. During the fitting procedure a constant value was used to account for the cross-calibration variations between the three EPIC detectors \citep[see][and the references therein]{2014A&A...564A..75R}. The EPIC X-ray spectra of EY Cyg are shown in the Fig.~\ref{fig:2}-bottom panel together with the best-fit model. The corresponding spectral parameters are given in Table~\ref{tab:2}.

\begin{table}[width=\linewidth,cols=4,pos=t]
\caption{Spectral Parameters of the best fit model to the EY Cyg Spectra.}
\label{tab:2}
\begin{tabular*}{\tblwidth}{@{} LLL@{} } 
\toprule

		\textbf{Model} & \textbf{Component} & \textbf{Value}\\
		\hline
		tbabs &  ${N_{H}}(\times 10^{21}cm^{-2})$ &  0.4$^{+0.2}_{-0.1}$ \\
		
		constant & & 	0.98$^{+0.03}_{-0.03}$ \\
 		mekal &  $kT(KeV)$ &  0.1$^{+0.01}_{-0.02}$ \\
 		&  $K _{MEKAL}(\times 10^{-5})$ &  2.7$^{+1.3}_{-1.3}$\\
		&  $F_{MKL} (\times 10^{-14} erg s^{-1} cm ^{-2})$ & 5.6$^{+3.2}_{-2.9}$\\
		&  $L_{X,MKL} (\times 10^{30} erg s^{-1})$ & 2.7$^{+1.5}_{-1.4}$ \\
		
		cevmkl &  $kT _{max} (keV)$ &  14.9$^{+3.3}_{-2.2}$\\
		&  $K _{CEVMKL} (\times 10^{-3})$  &  1.6$^{+0.2}_{-0.2}$\\
		&  $alpha$  &  1.2$^{+0.2}_{-0.2}$\\
		&  $F_{CEV} (\times 10^{-12} erg s^{-1} cm ^{-2})$ & 1.9$^{+0.1}_{-0.1}$\\
		&  $L_{X,CEV} (\times 10^{31} erg s^{-1})$ & 9.2$^{+0.5}_{-0.5}$ \\
       $\chi_{\nu}^{2}$ (dof) & & 1.04 (527)  \\
		 
\bottomrule
\end{tabular*}

\end{table}

In nonmagnetic CVs, the accreting material ($KT_{\rm max} \sim$ 6--55 keV) should be cooled before settling onto the white dwarf surface through the boundary layer \citep[see][]{Balman2015}. The structure of the boundary layer is not fully understood and the accretion flow structure in the inner disks of dwarf novae is non-standard (see the introduction). However, we can discuss the general characteristics of the X-ray emission from the source. The material is expected to cover a temperature distribution of hot optically thin cooling gas flow in collisional equilibrium \citep{2003ApJ...586L..77M,2005ApJ...626..396P,2005MNRAS.357..626B,2008ApJ...680..695O,2006MNRAS.372..450G,2011ApJ...741...84B}. Hence, in a simple way, the spectra of such systems are well modeled with an isobaric cooling flow which is a multi-temperature distribution of plasma characterized by an emission measure following a power-law temperature dependency (dEM = $(T/T_{\rm max})^{\alpha-1}$d$T/T_{\rm max}$) like MKCFLOW and CEVMKL \citep[built from the MEKAL code][]{1985A&AS...62..197M,1996ApJ...456..766S}.

The spectral fitting was simultaneously performed for all three EPIC spectra using a \textit{tbabs} model \citep{2000ApJ...542..914W} to account for the absorption through the interstellar medium and a multi-temperature plasma emission model (CEVMKL) as expected from low accretion rate quiescent dwarf novae. To begin with the fitting process, we fixed the neutral hydrogen column density at $N_{\rm H}$ = 9.48 $\times$ 10$^{21}$ atoms cm$^{-2}$ reported by nhtot\footnote{\url{http://www.swift.ac.uk/analysis/nhtot/}} \citep{2013MNRAS.431..394W}. However, the fit yielded an unacceptable $\chi^{2}_{\nu}$ of $\sim$10. We then let $N_{\rm H}$ vary where the fit was improved significantly with a $\chi^{2}_{\nu}$ $>$ 2. As shown in Fig.~\ref{fig:2}-top, the model was not able to fit the 3--4 sigma soft excess at energies below 0.5 keV. In order to fit the soft X-ray excess of the spectra, in our first attempt, we added a blackbody component to the composite model. Although the $\chi^{2}_{\nu}$ was 1.03, the fit yielded an unacceptable blackbody temperature of $\sim$200 keV. It is not in the range of temperatures predicted for soft X-ray emitting boundary layers \citep{popham1995accretion}. We, then, removed the blackbody component and added a power-law model instead. It yielded the fit parameters as  $\Gamma$ = 3.46$^{+0.04}_{-0.03}$ and norm = 4.5$^{+2.9}_{-2.5}$ $\times$ $10^{-5}$ with a $\chi^{2}_{\nu}$ of 0.99. However, the fit gave a column density $N_{\rm H}$ = 1.1 $^{+0.4}_{-0.3}$ $\times$ 10$^{21}$ which may not be a correct value as we discuss later in Sec.~\ref{sec:dis}. As the spectral parameters of both blackbody and power-low models were not physical, we used a single temperature plasma model \citep[\textit{MEKAL}][]{1985A&AS...62..197M,1995ApJ...438L.115L} to account for the soft excess. It perfectly fitted the excess resulting in a good $\chi^{2}_{\nu}$ 1.04 (see Table~\ref{tab:2}). In addition, the fit was relatively poor around the Fe K line region at 6-7 keV. Most probably the addition of the low-temperature \textit{Mekal} model permitted the CEVMKL model to shift slightly up in temperature and provide a better fit around the Fe K line region.


The best-fit model yields $kT_{\rm max}$ = 14.9$^{+3.3}_{-2.2}$ keV for the maximum plasma temperature, $\alpha$ = 1.2$^{+0.2}_{-0.2}$ for the power-law index of the temperature distribution and $K_{\rm CEVMKL}$ = 1.6$^{+0.2}_{-0.2}$ $\times$ 10$^{-3}$ for the CEVMKL normalization. \textit{MEKAL} temperature found to be $kT_{\rm MEKAL} = 0.1_{-0.02}^{+0.01}$ keV with a normalization $K_{\rm MEKAL}$ = 2.7$_{-1.3}^{+1.3}$ $\times$ 10$^{-5}$. The $N_{\rm H}$ was also calculated to be 0.4$^{+0.2}_{-0.1} \times 10^{21}$ cm$^{-2}$. All the errors were obtained at the 90\% confidence level. Finally, the acceptable composite model to fit the spectra was a combination of a \textit{MEKAL} and a CEVMKL together with an interstellar absorption (tbabs) which yielded the  $\chi^{2}_{\nu}$ of 1.04 (dof 527).

\section{Discussion}
\label{sec:dis}

In this work we present the X-ray temporal and spectral analysis of dwarf nova EY Cyg in quiescence using $\sim$45 ks of archival data obtained by \textit{XMM-Newton} observatory on 23 April 2007. A CEVMKL model was used to fit the hard X-rays emitted from the boundary layer as expected in low accretion rate quiescent dwarf novae. However, we detect a soft excess in the energies below 0.5 keV and it is not physically modeled by blackbody emission from a boundary layer. Therefore, instead of the blackbody model, we used a \textit{MEKAL} model to fit the soft X-rays which may emerge from the secondary estimated to be a K-type star \citep{1962ApJ...135..408K,1997MNRAS.287..271C}.

\begin{figure}
	\centering
	\includegraphics[width=1.0\linewidth]{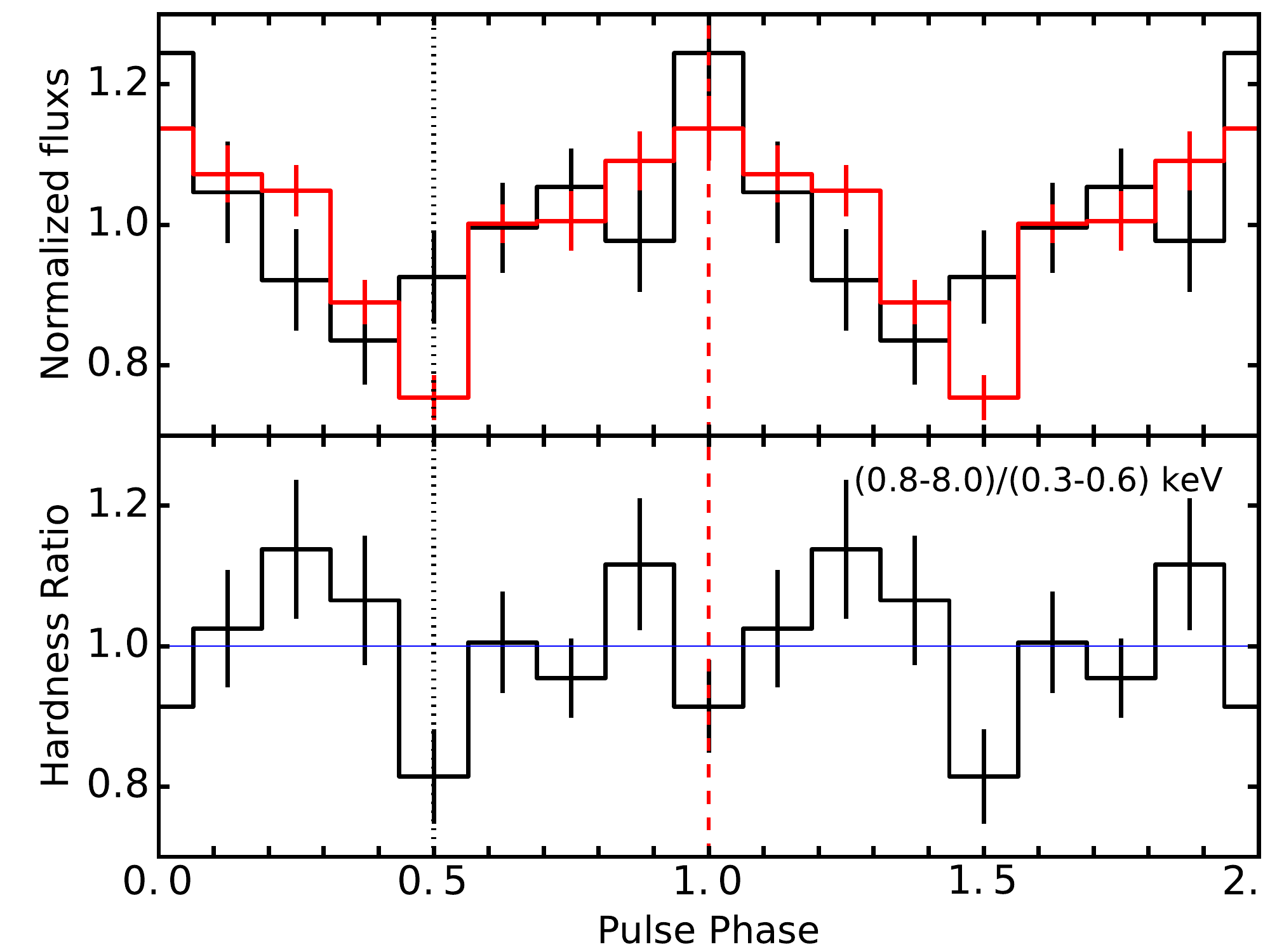}
	\caption{\textit{Top:} X-ray light curves of EY Cyg folded with the orbital period 11.0237976 h in two different energy bands 0.3--0.6 (in black) and 0.8--8 keV (in red) obtained with the \textit{XMM-Newton}/EPIC-pn. The fluxes are normalized by the mean flux.The vertical dashed and dotted lines are fixed at the location of main maximum and minimum, respectively. The ephemeris of $T_{0}$ = 2449255.3260(9)  \citep{2007A&A...462.1069E} was considered as the epoch for extracting the folded light curves. \textit{Bottom panel}: Hardness ratio of EY Cyg folded light curves in the energy bands 0.8--8.0/0.3--0.6 keV. The hardness ratio of unity is shown by the blue line.}
	\label{fig:3}
\end{figure}

Numerous observations provided by the early generation of X-ray space observatories such as Einstein, ROSAT and, EXOSAT have shown that all late-type main sequence stars are powerful X-ray emitters \citep{1981ApJ...245..163V,1981ApJ...248..279P,1989A&ARv...1..177P,1998A&AS..131..197M}. Using a photometric and low resolution spectroscopic analysis, \citet{1998A&AS..131..197M} investigated the spectral classification and X-ray luminosity of about 46 selected stars of type F, G and K observed by ROSAT in its all-sky survey. The mean X-ray luminosity they found for their samples was in the range of 10$^{29}$--10$^{30}$ erg s$^{-1}$. In the Table~\ref{tab:3}, we summarized the information of four K0-type stars they selected. According to this work and another analysis done by \citet{1992A&A...264L..31G} the X-ray luminosity of K-type stars can be roughly estimated to be around 10$^{30}$ erg s$^{-1}$. 
In addition, K stars, like other late type stars, are a source of thermal emission in which the coronal plasma temperature is in the range  of 10$^{6}$--10$^{7}$ K \cite{1989A&ARv...1..177P,2012A&A...543A..84R}. The spectral parameters of the \textit{MEKAL} model yields a temperature of $kT_{\rm MEKAL} = 0.1_{-0.01}^{+0.02}$ and an X-ray flux of $F_{\rm sec}$ = 5.6$^{+3.2}_{-2.9}$ $\times$ 10$^{-14}$ erg s$^{-1}$ cm$^{-2}$ in the energy band of 0.1--50 keV which gives an X-ray luminosity of $L_{\rm X,sec}$ = 1.3$^{+0.8}_{-0.7}$ $\times$ 10$^{30}$ erg s$^{-1}$. Both the temperature and X-ray luminosity are consistent with a K-type secondary star parameters. 

In order to make sure that the origin of the soft X-rays is the secondary star rather than the boundary layer, the investigation of the phase variations of the binary components, primary and secondary, was considered. For this, we separate the light curves regarding their energy range into two light curves with different energy bands of 0.3-6.0 keV and 0.8-8.0 keV. Using the epoch of the ephemeris used by \cite{2007A&A...462.1069E}, we again folded the light curves at the orbital period of the system calculated with radial velocity variations (Fig.~\ref{fig:3}-upper panel). Then, we plotted the hardness ratio of the two folded light curves with different energy bands (Fig.~\ref{fig:3}-lower panel). As clearly seen in the plots, there is a phase shift in the minima. To get more accurate value for the phase shifts, we applied a Cross Correlation Function (CCF) to the signals and obtained a phase shift $\Delta\phi$=0.1. The phase-shifted minima indicates that the origins of the X-ray modulations for the soft-band light curve (0.3--0.6 keV) and the hard-band light curve (0.8--8.0 keV) are different. The two average light curves indicate two sinusoidals that are shifted by about a phase of 0.1. Given the low inclination angle of the system that is 14, the small phase difference of the different components on the binary plane can yield the 0.1 phase shift. One being on the disk, the other at the secondary. It is not clear if the emitting region on the secondary is  coronal or the heated surface close to the L1 region. 
Our results are also consistent with the standard expectations which states that the boundary layer in quiescent state is optically thin radiating hard X-rays rather than soft X-rays, in general.

\begin{table*}
\caption{The X-Ray Luminosity of K0 Spectral Type Stars.}
\label{tab:3}
\begin{tabular*}{\tblwidth}{@{} LLLL@{} } 
\toprule
Name/ROSAT Name & Spectral Type & Distance & L$_{X}$\\
\midrule
		
RXJ 0540.0-5343 & K0V &  164$^{+24}_{-18}$ & 1.57$^{+1.9}_{-1.0} (\times 10^{30} erg s^{-1})$\\
RXJ 0548.0-6241 & K0/1V &  170$^{+34}_{-22}$ & 1.6$^{+2.3}_{-1.0} (\times 10^{30} erg s^{-1})$\\
RXJ 1429.4-0049 & K0V &  87$^{+6}_{-6}$ & 1.2$^{+0.5}_{-0.4} (\times 10^{30} erg s^{-1})$\\
RXJ 1433.3-0126 & K0/1V &  506$^{+116}_{-75}$ & 12.3$^{+19.2}_{-8.0} (\times 10^{30} erg s^{-1})$\\
\bottomrule
\end{tabular*}

\end{table*}

According to $nhtot$ task that calculates the neutral hydrogen column density using Gamma-ray burst data \citep{2013MNRAS.431..394W}, the column density for EY Cyg is 9.48 $\times$ 10$^{21}$ atoms cm$^{-2}$. In our fit, $N_{\rm H}$ was calculated to be 0.4$^{+0.2}_{-0.1} \times 10^{21}cm^{-2}$. Using the relation between the optical reddening and the hydrogen column density $N_{\rm H}$ = 5 $\times$ 10$^{21}$ $E(V-B)$\footnote{\url{https://heasarc.nasa.gov/Tools/w3nh_help.html}}, our $N_{\rm H}$ gives an optical reddening of 0.08 mag. The interstellar reddening for the source is not determined.
However, \citet{2004AJ....128.1795S} assumes a value of 0.1 for their disk calculations which is consistent with the UV nature of the source. We independently calculate the $N_{\rm H}$ and the E(B-V) from that and we confirm their assumption.

For EY Cyg, the unabsorbed X-ray flux is calculated to be 1.9$^{+0.1}_{-0.1}$ $\times$ 10$^{-12}$ erg s$^{-1}$ cm$^{-2}$ in energy rang of 0.1--50 keV. To find the X-ray flux in the wide energy range, we extrapolated the spectral model to 50 keV. The X-ray flux then can be translated to an X-ray luminosity of 4.6$^{+0.2}_{-0.2}$ $\times$ 10$^{31}$ erg s$^{-1}$ at the distance $\sim$637 pc. In addition, using the equation $L_{\rm acc}$ = $G\dot{M}M_{\rm WD}$/$R_{\rm WD}$ and by taking $M_{\rm WD}$ = 1.1 $M_{\odot}$ and $R_{\rm WD}$ = 6 $\times$ 10$^{8}$ \citep{1972ApJ...175..417N}, we obtained a mass accretion rate estimation for EY Cyg as 6.0$^{+0.3}_{-0.3}$ $\times$ 10$^{-12}$ $M_{\odot}$ yr$^{-1}$ which is expected for a source in a quiescent state. \citet{2004AJ....128.1795S} derived an accretion rate of 10$^{-10}$ $M_{\odot}$ yr$^{-1}$  using the far-UV spectra by applying a standard accretion disk model and a single-temperature WD emission model.

We find that the mass accretion rate calculated from X-rays is about a factor of 50 less than the UV and optical wavelengths.  
We caution that a distance of 450 pc were derived from the UV analysis which is not consistent with the GAIA result and thus the discrepancy between the UV and the X-ray accretion rates can be more. This can be well expected if the X-ray emitting region is composed of advective hot flows; ADAF-like accretion flows. Advection characteristics makes the flow region less luminous than what is expected from the UV and optical emitting regions on the disk since the flow retains the energy instead of radiating it \citep[see the detailed discussion in ][]{balman2014swift}. Thus, we find an efficiency factor $\epsilon\sim0.05$ in the X-rays. In addition, an efficiency range of 0.01-0.001 was calculated for nova-like systems by \citet{balman2014swift} \citep[see also ][]{kuulkers2006compact}.


\section{Summary and Conclusions}

This paper presents the X-ray analysis of $\sim$45 ks \textit{XMM-Newton} EPIC observations of poorly studied dwarf nova EY Cyg during the quiescent state. The EPIC spectra of the source are well fitted with an interstellar medium absorption model (\textit{tbabs}) together with a multi-temperature thermal plasma emission model (CEVMKL) and a single temperature thermal plasma model (\textit{MEKAL}). The maximum temperature of $\sim$15 keV was found for the source which is in the temperature range of nonmagnetic CVs \citep{Balman2015}. The unabsorbed X-ray flux of the source has been calculated to be 1.8--2.0 $\times$ 10$^{-12}$ erg s$^{-1}$ cm$^{-2}$ in the 0.1--50 keV which translates to an unabsorbed X-ray luminosity of 8.7--9.7 $\times$ 10$^{31}$ erg s$^{-1}$. The system has an accretion rate of 5.7--6.3 $\times$ 10$^{-12}$ $M_{\odot}$ yr$^{-1}$. We find that the X-ray emitting region shows advective hotflow characteristics and it is underluminous with an  efficiency of emission in this region about 0.01. We note that we expect the WD to be somewhat heated as a result of the advective heating from the inner ADAF-like flow. An $N_{\rm H}$ of 0.03--0.6 $\times$ 10$^{22}$ cm$^{-2}$ was derived in the X-ray spectral fitting that can be used to calculate an optical reddening of 0.08. It is in accordance with the low reddening of E(B-V) = 0.1 that was found consistent with the UV modeling \citep{2004AJ....128.1795S}. In addition, we detect significant (3-4 sigma) soft excess which we model with a thermal plasma (\textit{MEKAL}) model. The blackbody or power-law models do not yield physical parameters. We attribute this second component in the spectrum of EY Cyg as originating from the secondary star since the spectral parameters (temperature and luminosity) are consistent with K-type stars. According to the previous studies, the secondary star in this system was estimated to be a K-type star \citep{1962ApJ...135..408K,1997MNRAS.287..271C}. Therefore, based on the X-ray results, we also highly suggest that the nature of the secondary is a K-type star.

\section*{Acknowledgments}
AN and \c{S}B thanks to anonymous referees for careful reading of the manuscript. AN acknowledges 
partial support from T\"UBITAK, The Scientific and Technological Research Council of Turkey, through project 114F351.

%

\printcredits

\bibliographystyle{model2-names}

\bibliography{refs.bib}

\begin{thebibliography}{47}
\expandafter\ifx\csname natexlab\endcsname\relax\def\natexlab#1{#1}\fi
\providecommand{\url}[1]{\texttt{#1}}
\providecommand{\href}[2]{#2}
\providecommand{\path}[1]{#1}
\providecommand{\DOIprefix}{doi:}
\providecommand{\ArXivprefix}{arXiv:}
\providecommand{\URLprefix}{URL: }
\providecommand{\Pubmedprefix}{pmid:}
\providecommand{\doi}[1]{\href{http://dx.doi.org/#1}{\path{#1}}}
\providecommand{\Pubmed}[1]{\href{pmid:#1}{\path{#1}}}
\providecommand{\bibinfo}[2]{#2}
\ifx\xfnm\relax \def\xfnm[#1]{\unskip,\space#1}\fi
\bibitem[{{Arnaud}(1996)}]{1996ASPC..101...17A}
\bibinfo{author}{{Arnaud}, K.A.}, \bibinfo{year}{1996}.
\newblock \bibinfo{title}{{XSPEC: The First Ten Years}}, in:
  \bibinfo{editor}{{Jacoby}, G.H.}, \bibinfo{editor}{{Barnes}, J.} (Eds.),
  \bibinfo{booktitle}{Astronomical Data Analysis Software and Systems V}, pp.
  \bibinfo{pages}{17--20}.
\bibitem[{Balman(2014)}]{Balman2014}
\bibinfo{author}{Balman, S.}, \bibinfo{year}{2014}.
\newblock \bibinfo{title}{Pre-outburst chandra observations of the recurrent
  nova t pyxidis}.
\newblock \bibinfo{journal}{\aap} \bibinfo{volume}{572}, \bibinfo{pages}{A114}.
\bibitem[{Balman(2015)}]{Balman2015}
\bibinfo{author}{Balman, S.}, \bibinfo{year}{2015}.
\newblock \bibinfo{title}{{Inner Disk Structure of Dwarf Novae in the Light of
  X-Ray Observations}}.
\newblock \bibinfo{journal}{Acta Polytechnica CTU Proceedings}
  \bibinfo{volume}{2}, \bibinfo{pages}{116--122}.
\bibitem[{{Balman}(2019)}]{Balman2019}
\bibinfo{author}{{Balman}, {\c{S}}.}, \bibinfo{year}{2019}.
\newblock \bibinfo{title}{{Disk structure of cataclysmic variables and
  broadband noise characteristics in comparison with XRBs}}.
\newblock \bibinfo{journal}{Astronomische Nachrichten} \bibinfo{volume}{340},
  \bibinfo{pages}{296--301}.
\newblock \DOIprefix\doi{10.1002/asna.201913613}.
\bibitem[{Balman et~al.(2014)Balman, Godon and Sion}]{balman2014swift}
\bibinfo{author}{Balman, S.}, \bibinfo{author}{Godon, P.},
  \bibinfo{author}{Sion, E.M.}, \bibinfo{year}{2014}.
\newblock \bibinfo{title}{Swift x-ray telescope observations of the nova-like
  cataclysmic variables mv lyr, bz cam, and v592 cas}.
\newblock \bibinfo{journal}{\apj} \bibinfo{volume}{794}, \bibinfo{pages}{84}.
\bibitem[{{Balman} et~al.(2011){Balman}, {Godon}, {Sion}, {Ness}, {Schlegel},
  {Barrett} and {Szkody}}]{2011ApJ...741...84B}
\bibinfo{author}{{Balman}, {\c S}.}, \bibinfo{author}{{Godon}, P.},
  \bibinfo{author}{{Sion}, E.M.}, \bibinfo{author}{{Ness}, J.U.},
  \bibinfo{author}{{Schlegel}, E.}, \bibinfo{author}{{Barrett}, P.E.},
  \bibinfo{author}{{Szkody}, P.}, \bibinfo{year}{2011}.
\newblock \bibinfo{title}{{XMM-Newton Observations of the Dwarf Nova RU Peg in
  Quiescence: Probe of the Boundary Layer}}.
\newblock \bibinfo{journal}{\apj} \bibinfo{volume}{741}, \bibinfo{pages}{84}.
\newblock \DOIprefix\doi{10.1088/0004-637X/741/2/84},
  \href{http://arxiv.org/abs/1108.2662}{\tt arXiv:1108.2662}.
\bibitem[{{Balman} and {Revnivtsev}(2012)}]{Balman2012}
\bibinfo{author}{{Balman}, {\c S}.}, \bibinfo{author}{{Revnivtsev}, M.},
  \bibinfo{year}{2012}.
\newblock \bibinfo{title}{{X-ray variations in the inner accretion flow of
  dwarf novae}}.
\newblock \bibinfo{journal}{\aap} \bibinfo{volume}{546}, \bibinfo{pages}{A112}.
\newblock \DOIprefix\doi{10.1051/0004-6361/201219469},
  \href{http://arxiv.org/abs/1208.5940}{\tt arXiv:1208.5940}.
\bibitem[{{Baskill} et~al.(2005){Baskill}, {Wheatley} and
  {Osborne}}]{2005MNRAS.357..626B}
\bibinfo{author}{{Baskill}, D.S.}, \bibinfo{author}{{Wheatley}, P.J.},
  \bibinfo{author}{{Osborne}, J.P.}, \bibinfo{year}{2005}.
\newblock \bibinfo{title}{{The complete set of ASCA X-ray observations of
  non-magnetic cataclysmic variables}}.
\newblock \bibinfo{journal}{\mnras} \bibinfo{volume}{357},
  \bibinfo{pages}{626--644}.
\newblock \DOIprefix\doi{10.1111/j.1365-2966.2005.08677.x},
  \href{http://arxiv.org/abs/astro-ph/0502317}{\tt arXiv:astro-ph/0502317}.
\bibitem[{{Connon Smith} et~al.(1997){Connon Smith}, {Sarna}, {Catalan} and
  {Jones}}]{1997MNRAS.287..271C}
\bibinfo{author}{{Connon Smith}, R.}, \bibinfo{author}{{Sarna}, M.J.},
  \bibinfo{author}{{Catalan}, M.S.}, \bibinfo{author}{{Jones}, D.H.P.},
  \bibinfo{year}{1997}.
\newblock \bibinfo{title}{{The 8190-Angstroms sodium doublet in cataclysmic
  variables - IV. A survey of 22 objects}}.
\newblock \bibinfo{journal}{\mnras} \bibinfo{volume}{287},
  \bibinfo{pages}{271--286}.
\newblock \DOIprefix\doi{10.1093/mnras/287.2.271}.
\bibitem[{{Costero} et~al.(1998){Costero}, {Echevarria} and
  {Pineda}}]{1998AAS...192.8206C}
\bibinfo{author}{{Costero}, R.}, \bibinfo{author}{{Echevarria}, J.},
  \bibinfo{author}{{Pineda}, L.}, \bibinfo{year}{1998}.
\newblock \bibinfo{title}{{EY Cygni: A Test Case for low inclination
  cataclysmic binaries}}, in: \bibinfo{booktitle}{American Astronomical Society
  Meeting Abstracts \#192}, p. \bibinfo{pages}{1156}.
\bibitem[{{den Herder} et~al.(2001){den Herder}, {Brinkman}, {Kahn},
  {Branduardi-Raymont}, {Thomsen}, {Aarts}, {Audard}, {Bixler}, {den Boggende},
  {Cottam}, {Decker}, {Dubbeldam}, {Erd}, {Goulooze}, {G{\"u}del}, {Guttridge},
  {Hailey}, {Janabi}, {Kaastra}, {de Korte}, {van Leeuwen}, {Mauche},
  {McCalden}, {Mewe}, {Naber}, {Paerels}, {Peterson}, {Rasmussen}, {Rees},
  {Sakelliou}, {Sako}, {Spodek}, {Stern}, {Tamura}, {Tandy}, {de Vries},
  {Welch} and {Zehnder}}]{2001A&A...365L...7D}
\bibinfo{author}{{den Herder}, J.W.}, \bibinfo{author}{{Brinkman}, A.C.},
  \bibinfo{author}{{Kahn}, S.M.}, \bibinfo{author}{{Branduardi-Raymont}, G.},
  \bibinfo{author}{{Thomsen}, K.}, \bibinfo{author}{{Aarts}, H.},
  \bibinfo{author}{{Audard}, M.}, \bibinfo{author}{{Bixler}, J.V.},
  \bibinfo{author}{{den Boggende}, A.J.}, \bibinfo{author}{{Cottam}, J.},
  \bibinfo{author}{{Decker}, T.}, \bibinfo{author}{{Dubbeldam}, L.},
  \bibinfo{author}{{Erd}, C.}, \bibinfo{author}{{Goulooze}, H.},
  \bibinfo{author}{{G{\"u}del}, M.}, \bibinfo{author}{{Guttridge}, P.},
  \bibinfo{author}{{Hailey}, C.J.}, \bibinfo{author}{{Janabi}, K.A.},
  \bibinfo{author}{{Kaastra}, J.S.}, \bibinfo{author}{{de Korte}, P.A.J.},
  \bibinfo{author}{{van Leeuwen}, B.J.}, \bibinfo{author}{{Mauche}, C.},
  \bibinfo{author}{{McCalden}, A.J.}, \bibinfo{author}{{Mewe}, R.},
  \bibinfo{author}{{Naber}, A.}, \bibinfo{author}{{Paerels}, F.B.},
  \bibinfo{author}{{Peterson}, J.R.}, \bibinfo{author}{{Rasmussen}, A.P.},
  \bibinfo{author}{{Rees}, K.}, \bibinfo{author}{{Sakelliou}, I.},
  \bibinfo{author}{{Sako}, M.}, \bibinfo{author}{{Spodek}, J.},
  \bibinfo{author}{{Stern}, M.}, \bibinfo{author}{{Tamura}, T.},
  \bibinfo{author}{{Tandy}, J.}, \bibinfo{author}{{de Vries}, C.P.},
  \bibinfo{author}{{Welch}, S.}, \bibinfo{author}{{Zehnder}, A.},
  \bibinfo{year}{2001}.
\newblock \bibinfo{title}{{The Reflection Grating Spectrometer on board
  XMM-Newton}}.
\newblock \bibinfo{journal}{\aap} \bibinfo{volume}{365},
  \bibinfo{pages}{L7--L17}.
\newblock \DOIprefix\doi{10.1051/0004-6361:20000058}.
\bibitem[{{Echevarr{\'{\i}}a} et~al.(2007){Echevarr{\'{\i}}a}, {Michel},
  {Costero} and {Zharikov}}]{2007A&A...462.1069E}
\bibinfo{author}{{Echevarr{\'{\i}}a}, J.}, \bibinfo{author}{{Michel}, R.},
  \bibinfo{author}{{Costero}, R.}, \bibinfo{author}{{Zharikov}, S.},
  \bibinfo{year}{2007}.
\newblock \bibinfo{title}{{Determination of the basic parameters of the dwarf
  nova EY Cygni}}.
\newblock \bibinfo{journal}{\aap} \bibinfo{volume}{462},
  \bibinfo{pages}{1069--1080}.
\newblock \DOIprefix\doi{10.1051/0004-6361:20052906},
  \href{http://arxiv.org/abs/astro-ph/0611267}{\tt arXiv:astro-ph/0611267}.
\bibitem[{{Gudel}(1992)}]{1992A&A...264L..31G}
\bibinfo{author}{{Gudel}, M.}, \bibinfo{year}{1992}.
\newblock \bibinfo{title}{{Radio and X-ray emission from main-sequence K
  stars}}.
\newblock \bibinfo{journal}{\aap} \bibinfo{volume}{264},
  \bibinfo{pages}{L31--L34}.
\bibitem[{{G{\"u}ver} et~al.(2006){G{\"u}ver}, {Uluyaz{\i}}, {{\"O}zkan} and
  {G{\"o}{\v g}{\"u}{\c s}}}]{2006MNRAS.372..450G}
\bibinfo{author}{{G{\"u}ver}, T.}, \bibinfo{author}{{Uluyaz{\i}}, C.},
  \bibinfo{author}{{{\"O}zkan}, M.T.}, \bibinfo{author}{{G{\"o}{\v g}{\"u}{\c
  s}}, E.}, \bibinfo{year}{2006}.
\newblock \bibinfo{title}{{X-ray spectral variations of U Gem from quiescence
  to outburst}}.
\newblock \bibinfo{journal}{\mnras} \bibinfo{volume}{372},
  \bibinfo{pages}{450--456}.
\newblock \DOIprefix\doi{10.1111/j.1365-2966.2006.10896.x},
  \href{http://arxiv.org/abs/astro-ph/0608084}{\tt arXiv:astro-ph/0608084}.
\bibitem[{{Harrison}(2018)}]{Harrison2018ApJ}
\bibinfo{author}{{Harrison}, T.E.}, \bibinfo{year}{2018}.
\newblock \bibinfo{title}{{The Identification of Hydrogen-deficient Cataclysmic
  Variable Donor Stars}}.
\newblock \bibinfo{journal}{\apj} \bibinfo{volume}{861}, \bibinfo{pages}{102}.
\newblock \DOIprefix\doi{10.3847/1538-4357/aacbd9},
  \href{http://arxiv.org/abs/1806.04612}{\tt arXiv:1806.04612}.
\bibitem[{{Jansen} et~al.(2001){Jansen}, {Lumb}, {Altieri}, {Clavel}, {Ehle},
  {Erd}, {Gabriel}, {Guainazzi}, {Gondoin}, {Much}, {Munoz}, {Santos},
  {Schartel}, {Texier} and {Vacanti}}]{2001A&A...365L...1J}
\bibinfo{author}{{Jansen}, F.}, \bibinfo{author}{{Lumb}, D.},
  \bibinfo{author}{{Altieri}, B.}, \bibinfo{author}{{Clavel}, J.},
  \bibinfo{author}{{Ehle}, M.}, \bibinfo{author}{{Erd}, C.},
  \bibinfo{author}{{Gabriel}, C.}, \bibinfo{author}{{Guainazzi}, M.},
  \bibinfo{author}{{Gondoin}, P.}, \bibinfo{author}{{Much}, R.},
  \bibinfo{author}{{Munoz}, R.}, \bibinfo{author}{{Santos}, M.},
  \bibinfo{author}{{Schartel}, N.}, \bibinfo{author}{{Texier}, D.},
  \bibinfo{author}{{Vacanti}, G.}, \bibinfo{year}{2001}.
\newblock \bibinfo{title}{{XMM-Newton observatory. I. The spacecraft and
  operations}}.
\newblock \bibinfo{journal}{\aap} \bibinfo{volume}{365},
  \bibinfo{pages}{L1--L6}.
\newblock \DOIprefix\doi{10.1051/0004-6361:20000036}.
\bibitem[{{Kraft}(1962)}]{1962ApJ...135..408K}
\bibinfo{author}{{Kraft}, R.P.}, \bibinfo{year}{1962}.
\newblock \bibinfo{title}{{Binary Stars among Cataclysmic Variables. I. U
  Geminorum Stars (dwarf Novae).}}
\newblock \bibinfo{journal}{\apj} \bibinfo{volume}{135},
  \bibinfo{pages}{408--423}.
\newblock \DOIprefix\doi{10.1086/147280}.
\bibitem[{Kuulkers et~al.(2006)Kuulkers, Norton, Schwope and
  Warner}]{kuulkers2006compact}
\bibinfo{author}{Kuulkers, E.}, \bibinfo{author}{Norton, A.},
  \bibinfo{author}{Schwope, A.}, \bibinfo{author}{Warner, B.},
  \bibinfo{year}{2006}.
\newblock \bibinfo{title}{X-rays from cataclysmic variables}.
\newblock \bibinfo{journal}{in Compact stellar X-ray sources, ed. W. Lewin \&
  M. van der Klis, Cambridge Astrophysics Series, Cambridge University Press,
  Cambridge,} \bibinfo{volume}{39}, \bibinfo{pages}{421--460}.
\bibitem[{{Liedahl} et~al.(1995){Liedahl}, {Osterheld} and
  {Goldstein}}]{1995ApJ...438L.115L}
\bibinfo{author}{{Liedahl}, D.A.}, \bibinfo{author}{{Osterheld}, A.L.},
  \bibinfo{author}{{Goldstein}, W.H.}, \bibinfo{year}{1995}.
\newblock \bibinfo{title}{{New calculations of Fe L-shell X-ray spectra in
  high-temperature plasmas}}.
\newblock \bibinfo{journal}{\apjl} \bibinfo{volume}{438},
  \bibinfo{pages}{L115--L118}.
\newblock \DOIprefix\doi{10.1086/187729}.
\bibitem[{{Lynden-Bell} and {Pringle}(1974)}]{1974MNRAS.168..603L}
\bibinfo{author}{{Lynden-Bell}, D.}, \bibinfo{author}{{Pringle}, J.E.},
  \bibinfo{year}{1974}.
\newblock \bibinfo{title}{{The evolution of viscous discs and the origin of the
  nebular variables.}}
\newblock \bibinfo{journal}{\mnras} \bibinfo{volume}{168},
  \bibinfo{pages}{603--637}.
\newblock \DOIprefix\doi{10.1093/mnras/168.3.603}.
\bibitem[{{Mason} et~al.(2001){Mason}, {Breeveld}, {Much}, {Carter}, {Cordova},
  {Cropper}, {Fordham}, {Huckle}, {Ho}, {Kawakami}, {Kennea}, {Kennedy},
  {Mittaz}, {Pandel}, {Priedhorsky}, {Sasseen}, {Shirey}, {Smith} and
  {Vreux}}]{2001A&A...365L..36M}
\bibinfo{author}{{Mason}, K.O.}, \bibinfo{author}{{Breeveld}, A.},
  \bibinfo{author}{{Much}, R.}, \bibinfo{author}{{Carter}, M.},
  \bibinfo{author}{{Cordova}, F.A.}, \bibinfo{author}{{Cropper}, M.S.},
  \bibinfo{author}{{Fordham}, J.}, \bibinfo{author}{{Huckle}, H.},
  \bibinfo{author}{{Ho}, C.}, \bibinfo{author}{{Kawakami}, H.},
  \bibinfo{author}{{Kennea}, J.}, \bibinfo{author}{{Kennedy}, T.},
  \bibinfo{author}{{Mittaz}, J.}, \bibinfo{author}{{Pandel}, D.},
  \bibinfo{author}{{Priedhorsky}, W.C.}, \bibinfo{author}{{Sasseen}, T.},
  \bibinfo{author}{{Shirey}, R.}, \bibinfo{author}{{Smith}, P.},
  \bibinfo{author}{{Vreux}, J.M.}, \bibinfo{year}{2001}.
\newblock \bibinfo{title}{{The XMM-Newton optical/UV monitor telescope}}.
\newblock \bibinfo{journal}{\aap} \bibinfo{volume}{365},
  \bibinfo{pages}{L36--L44}.
\newblock \DOIprefix\doi{10.1051/0004-6361:20000044},
  \href{http://arxiv.org/abs/astro-ph/0011216}{\tt arXiv:astro-ph/0011216}.
\bibitem[{{Metanomski} et~al.(1998){Metanomski}, {Pasquini}, {Krautter},
  {Cutispoto} and {Fleming}}]{1998A&AS..131..197M}
\bibinfo{author}{{Metanomski}, A.D.F.}, \bibinfo{author}{{Pasquini}, L.},
  \bibinfo{author}{{Krautter}, J.}, \bibinfo{author}{{Cutispoto}, G.},
  \bibinfo{author}{{Fleming}, T.A.}, \bibinfo{year}{1998}.
\newblock \bibinfo{title}{{F, G and K stars in the ROSAT all-sky survey. I.
  Photometry}}.
\newblock \bibinfo{journal}{\aaps} \bibinfo{volume}{131},
  \bibinfo{pages}{197--208}.
\newblock \DOIprefix\doi{10.1051/aas:1998431}.
\bibitem[{{Mewe} et~al.(1985){Mewe}, {Gronenschild} and {van den
  Oord}}]{1985A&AS...62..197M}
\bibinfo{author}{{Mewe}, R.}, \bibinfo{author}{{Gronenschild}, E.H.B.M.},
  \bibinfo{author}{{van den Oord}, G.H.J.}, \bibinfo{year}{1985}.
\newblock \bibinfo{title}{{Calculated X-radiation from optically thin plasmas.
  V}}.
\newblock \bibinfo{journal}{\aaps} \bibinfo{volume}{62},
  \bibinfo{pages}{197--254}.
\bibitem[{{Mouchet} et~al.(2012){Mouchet}, {Bonnet-Bidaud} and {de
  Martino}}]{2012MmSAI..83..578M}
\bibinfo{author}{{Mouchet}, M.}, \bibinfo{author}{{Bonnet-Bidaud}, J.M.},
  \bibinfo{author}{{de Martino}, D.}, \bibinfo{year}{2012}.
\newblock \bibinfo{title}{{The X-ray emission of magnetic cataclysmic variables
  in the XMM-Newton era.}}
\newblock \bibinfo{journal}{\memsai} \bibinfo{volume}{83},
  \bibinfo{pages}{578}.
\newblock \href{http://arxiv.org/abs/1202.3594}{\tt arXiv:1202.3594}.
\bibitem[{Mukai(2017)}]{mukai2017x}
\bibinfo{author}{Mukai, K.}, \bibinfo{year}{2017}.
\newblock \bibinfo{title}{X-ray emissions from accreting white dwarfs: A
  review}.
\newblock \bibinfo{journal}{Publications of the Astronomical Society of the
  Pacific} \bibinfo{volume}{129}, \bibinfo{pages}{062001}.
\bibitem[{{Mukai} et~al.(2003){Mukai}, {Kinkhabwala}, {Peterson}, {Kahn} and
  {Paerels}}]{2003ApJ...586L..77M}
\bibinfo{author}{{Mukai}, K.}, \bibinfo{author}{{Kinkhabwala}, A.},
  \bibinfo{author}{{Peterson}, J.R.}, \bibinfo{author}{{Kahn}, S.M.},
  \bibinfo{author}{{Paerels}, F.}, \bibinfo{year}{2003}.
\newblock \bibinfo{title}{{Two Types of X-Ray Spectra in Cataclysmic
  Variables}}.
\newblock \bibinfo{journal}{\apjl} \bibinfo{volume}{586},
  \bibinfo{pages}{L77--L80}.
\newblock \DOIprefix\doi{10.1086/374583},
  \href{http://arxiv.org/abs/astro-ph/0301557}{\tt arXiv:astro-ph/0301557}.
\bibitem[{{Nauenberg}(1972)}]{1972ApJ...175..417N}
\bibinfo{author}{{Nauenberg}, M.}, \bibinfo{year}{1972}.
\newblock \bibinfo{title}{{Analytic Approximations to the Mass-Radius Relation
  and Energy of Zero-Temperature Stars}}.
\newblock \bibinfo{journal}{\apj} \bibinfo{volume}{175},
  \bibinfo{pages}{417--430}.
\newblock \DOIprefix\doi{10.1086/151568}.
\bibitem[{{Okada} et~al.(2008){Okada}, {Nakamura} and
  {Ishida}}]{2008ApJ...680..695O}
\bibinfo{author}{{Okada}, S.}, \bibinfo{author}{{Nakamura}, R.},
  \bibinfo{author}{{Ishida}, M.}, \bibinfo{year}{2008}.
\newblock \bibinfo{title}{{Chandra HETG Line Spectroscopy of the Nonmagnetic
  Cataclysmic Variable SS Cygni}}.
\newblock \bibinfo{journal}{\apj} \bibinfo{volume}{680},
  \bibinfo{pages}{695--704}.
\newblock \DOIprefix\doi{10.1086/587161},
  \href{http://arxiv.org/abs/0805.2724}{\tt arXiv:0805.2724}.
\bibitem[{{Orio} and {Ogelman}(1992)}]{1992IAUC.5680....2O}
\bibinfo{author}{{Orio}, M.}, \bibinfo{author}{{Ogelman}, H.},
  \bibinfo{year}{1992}.
\newblock \bibinfo{title}{{EY Cygni}}.
\newblock \bibinfo{journal}{International Astronomical Union Circular}
  \bibinfo{volume}{5680}, \bibinfo{pages}{2}.
\bibitem[{{Pallavicini}(1989)}]{1989A&ARv...1..177P}
\bibinfo{author}{{Pallavicini}, R.}, \bibinfo{year}{1989}.
\newblock \bibinfo{title}{{X-ray emission from stellar coronae}}.
\newblock \bibinfo{journal}{\aapr} \bibinfo{volume}{1},
  \bibinfo{pages}{177--207}.
\newblock \DOIprefix\doi{10.1007/BF00872715}.
\bibitem[{{Pallavicini} et~al.(1981){Pallavicini}, {Golub}, {Rosner}, {Vaiana},
  {Ayres} and {Linsky}}]{1981ApJ...248..279P}
\bibinfo{author}{{Pallavicini}, R.}, \bibinfo{author}{{Golub}, L.},
  \bibinfo{author}{{Rosner}, R.}, \bibinfo{author}{{Vaiana}, G.S.},
  \bibinfo{author}{{Ayres}, T.}, \bibinfo{author}{{Linsky}, J.L.},
  \bibinfo{year}{1981}.
\newblock \bibinfo{title}{{Relations among stellar X-ray emission observed from
  Einstein, stellar rotation and bolometric luminosity.}}
\newblock \bibinfo{journal}{\apj} \bibinfo{volume}{248},
  \bibinfo{pages}{279--290}.
\newblock \DOIprefix\doi{10.1086/159152}.
\bibitem[{{Pandel} et~al.(2005){Pandel}, {C{\'o}rdova}, {Mason} and
  {Priedhorsky}}]{2005ApJ...626..396P}
\bibinfo{author}{{Pandel}, D.}, \bibinfo{author}{{C{\'o}rdova}, F.A.},
  \bibinfo{author}{{Mason}, K.O.}, \bibinfo{author}{{Priedhorsky}, W.C.},
  \bibinfo{year}{2005}.
\newblock \bibinfo{title}{{X-Ray Observations of the Boundary Layer in Dwarf
  Novae at Low Accretion Rates}}.
\newblock \bibinfo{journal}{\apj} \bibinfo{volume}{626},
  \bibinfo{pages}{396--410}.
\newblock \DOIprefix\doi{10.1086/429983},
  \href{http://arxiv.org/abs/astro-ph/0503114}{\tt arXiv:astro-ph/0503114}.
\bibitem[{{Patterson} and {Raymond}(1985)}]{1985ApJ...292..535P}
\bibinfo{author}{{Patterson}, J.}, \bibinfo{author}{{Raymond}, J.C.},
  \bibinfo{year}{1985}.
\newblock \bibinfo{title}{{X-ray emission from cataclysmic variables with
  accretion disks. I - Hard X-rays. II - EUV/soft X-ray radiation}}.
\newblock \bibinfo{journal}{\apj} \bibinfo{volume}{292},
  \bibinfo{pages}{535--558}.
\newblock \DOIprefix\doi{10.1086/163187}.
\bibitem[{{Piening}(1978)}]{1978JAVSO...6...60P}
\bibinfo{author}{{Piening}, A.T.}, \bibinfo{year}{1978}.
\newblock \bibinfo{title}{{The U Geminorum star, EY Cygni, and a Probable
  Eclipsing Star, V839 Cygni}}.
\newblock \bibinfo{journal}{Journal of the American Association of Variable
  Star Observers (JAAVSO)} \bibinfo{volume}{6}, \bibinfo{pages}{60--61}.
\bibitem[{{Popham} and {Narayan}(1995)}]{popham1995accretion}
\bibinfo{author}{{Popham}, R.}, \bibinfo{author}{{Narayan}, R.},
  \bibinfo{year}{1995}.
\newblock \bibinfo{title}{{Accretion disk boundary layers in cataclysmic
  variables. 1: Optically thick boundary layers}}.
\newblock \bibinfo{journal}{\apj} \bibinfo{volume}{442},
  \bibinfo{pages}{337--357}.
\newblock \DOIprefix\doi{10.1086/175444}.
\bibitem[{{Read} et~al.(2014){Read}, {Guainazzi} and
  {Sembay}}]{2014A&A...564A..75R}
\bibinfo{author}{{Read}, A.M.}, \bibinfo{author}{{Guainazzi}, M.},
  \bibinfo{author}{{Sembay}, S.}, \bibinfo{year}{2014}.
\newblock \bibinfo{title}{{Cross-calibration of the XMM-Newton EPIC pn and MOS
  on-axis effective areas using 2XMM sources}}.
\newblock \bibinfo{journal}{\aap} \bibinfo{volume}{564}, \bibinfo{pages}{A75}.
\newblock \DOIprefix\doi{10.1051/0004-6361/201423422},
  \href{http://arxiv.org/abs/1403.3555}{\tt arXiv:1403.3555}.
\bibitem[{{Robrade} et~al.(2012){Robrade}, {Schmitt} and
  {Favata}}]{2012A&A...543A..84R}
\bibinfo{author}{{Robrade}, J.}, \bibinfo{author}{{Schmitt}, J.H.M.M.},
  \bibinfo{author}{{Favata}, F.}, \bibinfo{year}{2012}.
\newblock \bibinfo{title}{{Coronal activity cycles in nearby G and K stars.
  XMM-Newton monitoring of 61 Cygni and {\ensuremath{\alpha}} Centauri}}.
\newblock \bibinfo{journal}{\aap} \bibinfo{volume}{543}, \bibinfo{pages}{A84}.
\newblock \DOIprefix\doi{10.1051/0004-6361/201219046},
  \href{http://arxiv.org/abs/1205.3627}{\tt arXiv:1205.3627}.
\bibitem[{{Singh} et~al.(1996){Singh}, {White} and
  {Drake}}]{1996ApJ...456..766S}
\bibinfo{author}{{Singh}, K.P.}, \bibinfo{author}{{White}, N.E.},
  \bibinfo{author}{{Drake}, S.A.}, \bibinfo{year}{1996}.
\newblock \bibinfo{title}{{Corona(e) of AR Lacertae. I. The Temperature and
  Abundance Distribution}}.
\newblock \bibinfo{journal}{\apj} \bibinfo{volume}{456},
  \bibinfo{pages}{766--776}.
\newblock \DOIprefix\doi{10.1086/176695}.
\bibitem[{{Sion} et~al.(2004){Sion}, {Winter}, {Urban}, {Tovmassian},
  {Zharikov}, {G{\"a}nsicke} and {Orio}}]{2004AJ....128.1795S}
\bibinfo{author}{{Sion}, E.M.}, \bibinfo{author}{{Winter}, L.},
  \bibinfo{author}{{Urban}, J.A.}, \bibinfo{author}{{Tovmassian}, G.H.},
  \bibinfo{author}{{Zharikov}, S.}, \bibinfo{author}{{G{\"a}nsicke}, B.T.},
  \bibinfo{author}{{Orio}, M.}, \bibinfo{year}{2004}.
\newblock \bibinfo{title}{{Composite Accretion Disk and White Dwarf Photosphere
  Analyses of the FUSE and Hubble Space Telescope Observations of EY Cygni}}.
\newblock \bibinfo{journal}{\aj} \bibinfo{volume}{128},
  \bibinfo{pages}{1795--1801}.
\newblock \DOIprefix\doi{10.1086/423995},
  \href{http://arxiv.org/abs/astro-ph/0407534}{\tt arXiv:astro-ph/0407534}.
\bibitem[{{Str{\"u}der} et~al.(2001){Str{\"u}der}, {Briel}, {Dennerl},
  {Hartmann}, {Kendziorra}, {Meidinger}, {Pfeffermann}, {Reppin}, {Aschenbach},
  {Bornemann}, {Br{\"a}uninger}, {Burkert}, {Elender}, {Freyberg}, {Haberl},
  {Hartner}, {Heuschmann}, {Hippmann}, {Kastelic}, {Kemmer}, {Kettenring},
  {Kink}, {Krause}, {M{\"u}ller}, {Oppitz}, {Pietsch}, {Popp}, {Predehl},
  {Read}, {Stephan}, {St{\"o}tter}, {Tr{\"u}mper}, {Holl}, {Kemmer}, {Soltau},
  {St{\"o}tter}, {Weber}, {Weichert}, {von Zanthier}, {Carathanassis}, {Lutz},
  {Richter}, {Solc}, {B{\"o}ttcher}, {Kuster}, {Staubert}, {Abbey}, {Holland},
  {Turner}, {Balasini}, {Bignami}, {La Palombara}, {Villa}, {Buttler},
  {Gianini}, {Lain{\'e}}, {Lumb} and {Dhez}}]{2001A&A...365L..18S}
\bibinfo{author}{{Str{\"u}der}, L.}, \bibinfo{author}{{Briel}, U.},
  \bibinfo{author}{{Dennerl}, K.}, \bibinfo{author}{{Hartmann}, R.},
  \bibinfo{author}{{Kendziorra}, E.}, \bibinfo{author}{{Meidinger}, N.},
  \bibinfo{author}{{Pfeffermann}, E.}, \bibinfo{author}{{Reppin}, C.},
  \bibinfo{author}{{Aschenbach}, B.}, \bibinfo{author}{{Bornemann}, W.},
  \bibinfo{author}{{Br{\"a}uninger}, H.}, \bibinfo{author}{{Burkert}, W.},
  \bibinfo{author}{{Elender}, M.}, \bibinfo{author}{{Freyberg}, M.},
  \bibinfo{author}{{Haberl}, F.}, \bibinfo{author}{{Hartner}, G.},
  \bibinfo{author}{{Heuschmann}, F.}, \bibinfo{author}{{Hippmann}, H.},
  \bibinfo{author}{{Kastelic}, E.}, \bibinfo{author}{{Kemmer}, S.},
  \bibinfo{author}{{Kettenring}, G.}, \bibinfo{author}{{Kink}, W.},
  \bibinfo{author}{{Krause}, N.}, \bibinfo{author}{{M{\"u}ller}, S.},
  \bibinfo{author}{{Oppitz}, A.}, \bibinfo{author}{{Pietsch}, W.},
  \bibinfo{author}{{Popp}, M.}, \bibinfo{author}{{Predehl}, P.},
  \bibinfo{author}{{Read}, A.}, \bibinfo{author}{{Stephan}, K.H.},
  \bibinfo{author}{{St{\"o}tter}, D.}, \bibinfo{author}{{Tr{\"u}mper}, J.},
  \bibinfo{author}{{Holl}, P.}, \bibinfo{author}{{Kemmer}, J.},
  \bibinfo{author}{{Soltau}, H.}, \bibinfo{author}{{St{\"o}tter}, R.},
  \bibinfo{author}{{Weber}, U.}, \bibinfo{author}{{Weichert}, U.},
  \bibinfo{author}{{von Zanthier}, C.}, \bibinfo{author}{{Carathanassis}, D.},
  \bibinfo{author}{{Lutz}, G.}, \bibinfo{author}{{Richter}, R.H.},
  \bibinfo{author}{{Solc}, P.}, \bibinfo{author}{{B{\"o}ttcher}, H.},
  \bibinfo{author}{{Kuster}, M.}, \bibinfo{author}{{Staubert}, R.},
  \bibinfo{author}{{Abbey}, A.}, \bibinfo{author}{{Holland}, A.},
  \bibinfo{author}{{Turner}, M.}, \bibinfo{author}{{Balasini}, M.},
  \bibinfo{author}{{Bignami}, G.F.}, \bibinfo{author}{{La Palombara}, N.},
  \bibinfo{author}{{Villa}, G.}, \bibinfo{author}{{Buttler}, W.},
  \bibinfo{author}{{Gianini}, F.}, \bibinfo{author}{{Lain{\'e}}, R.},
  \bibinfo{author}{{Lumb}, D.}, \bibinfo{author}{{Dhez}, P.},
  \bibinfo{year}{2001}.
\newblock \bibinfo{title}{{The European Photon Imaging Camera on XMM-Newton:
  The pn-CCD camera}}.
\newblock \bibinfo{journal}{\aap} \bibinfo{volume}{365},
  \bibinfo{pages}{L18--L26}.
\newblock \DOIprefix\doi{10.1051/0004-6361:20000066}.
\bibitem[{{Tovmassian} et~al.(2002){Tovmassian}, {Orio}, {Zharikov},
  {Echevarr{\'{\i}}a}, {Costero} and {Michel}}]{2002AIPC..637...72T}
\bibinfo{author}{{Tovmassian}, G.}, \bibinfo{author}{{Orio}, M.},
  \bibinfo{author}{{Zharikov}, S.}, \bibinfo{author}{{Echevarr{\'{\i}}a}, J.},
  \bibinfo{author}{{Costero}, R.}, \bibinfo{author}{{Michel}, R.},
  \bibinfo{year}{2002}.
\newblock \bibinfo{title}{{Did EY Cyg go through a Nova explosion?}}, in:
  \bibinfo{editor}{{Hernanz}, M.}, \bibinfo{editor}{{Jos{\'e}}, J.} (Eds.),
  \bibinfo{booktitle}{Classical Nova Explosions}, pp. \bibinfo{pages}{72--76}.
\newblock \DOIprefix\doi{10.1063/1.1518181}.
\bibitem[{{Turner} et~al.(2001){Turner}, {Abbey}, {Arnaud}, {Balasini},
  {Barbera}, {Belsole}, {Bennie}, {Bernard}, {Bignami}, {Boer}, {Briel},
  {Butler}, {Cara}, {Chabaud}, {Cole}, {Collura}, {Conte}, {Cros}, {Denby},
  {Dhez}, {Di Coco}, {Dowson}, {Ferrando}, {Ghizzardi}, {Gianotti}, {Goodall},
  {Gretton}, {Griffiths}, {Hainaut}, {Hochedez}, {Holland}, {Jourdain},
  {Kendziorra}, {Lagostina}, {Laine}, {La Palombara}, {Lortholary}, {Lumb},
  {Marty}, {Molendi}, {Pigot}, {Poindron}, {Pounds}, {Reeves}, {Reppin},
  {Rothenflug}, {Salvetat}, {Sauvageot}, {Schmitt}, {Sembay}, {Short},
  {Spragg}, {Stephen}, {Str{\"u}der}, {Tiengo}, {Trifoglio}, {Tr{\"u}mper},
  {Vercellone}, {Vigroux}, {Villa}, {Ward}, {Whitehead} and
  {Zonca}}]{2001A&A...365L..27T}
\bibinfo{author}{{Turner}, M.J.L.}, \bibinfo{author}{{Abbey}, A.},
  \bibinfo{author}{{Arnaud}, M.}, \bibinfo{author}{{Balasini}, M.},
  \bibinfo{author}{{Barbera}, M.}, \bibinfo{author}{{Belsole}, E.},
  \bibinfo{author}{{Bennie}, P.J.}, \bibinfo{author}{{Bernard}, J.P.},
  \bibinfo{author}{{Bignami}, G.F.}, \bibinfo{author}{{Boer}, M.},
  \bibinfo{author}{{Briel}, U.}, \bibinfo{author}{{Butler}, I.},
  \bibinfo{author}{{Cara}, C.}, \bibinfo{author}{{Chabaud}, C.},
  \bibinfo{author}{{Cole}, R.}, \bibinfo{author}{{Collura}, A.},
  \bibinfo{author}{{Conte}, M.}, \bibinfo{author}{{Cros}, A.},
  \bibinfo{author}{{Denby}, M.}, \bibinfo{author}{{Dhez}, P.},
  \bibinfo{author}{{Di Coco}, G.}, \bibinfo{author}{{Dowson}, J.},
  \bibinfo{author}{{Ferrando}, P.}, \bibinfo{author}{{Ghizzardi}, S.},
  \bibinfo{author}{{Gianotti}, F.}, \bibinfo{author}{{Goodall}, C.V.},
  \bibinfo{author}{{Gretton}, L.}, \bibinfo{author}{{Griffiths}, R.G.},
  \bibinfo{author}{{Hainaut}, O.}, \bibinfo{author}{{Hochedez}, J.F.},
  \bibinfo{author}{{Holland}, A.D.}, \bibinfo{author}{{Jourdain}, E.},
  \bibinfo{author}{{Kendziorra}, E.}, \bibinfo{author}{{Lagostina}, A.},
  \bibinfo{author}{{Laine}, R.}, \bibinfo{author}{{La Palombara}, N.},
  \bibinfo{author}{{Lortholary}, M.}, \bibinfo{author}{{Lumb}, D.},
  \bibinfo{author}{{Marty}, P.}, \bibinfo{author}{{Molendi}, S.},
  \bibinfo{author}{{Pigot}, C.}, \bibinfo{author}{{Poindron}, E.},
  \bibinfo{author}{{Pounds}, K.A.}, \bibinfo{author}{{Reeves}, J.N.},
  \bibinfo{author}{{Reppin}, C.}, \bibinfo{author}{{Rothenflug}, R.},
  \bibinfo{author}{{Salvetat}, P.}, \bibinfo{author}{{Sauvageot}, J.L.},
  \bibinfo{author}{{Schmitt}, D.}, \bibinfo{author}{{Sembay}, S.},
  \bibinfo{author}{{Short}, A.D.T.}, \bibinfo{author}{{Spragg}, J.},
  \bibinfo{author}{{Stephen}, J.}, \bibinfo{author}{{Str{\"u}der}, L.},
  \bibinfo{author}{{Tiengo}, A.}, \bibinfo{author}{{Trifoglio}, M.},
  \bibinfo{author}{{Tr{\"u}mper}, J.}, \bibinfo{author}{{Vercellone}, S.},
  \bibinfo{author}{{Vigroux}, L.}, \bibinfo{author}{{Villa}, G.},
  \bibinfo{author}{{Ward}, M.J.}, \bibinfo{author}{{Whitehead}, S.},
  \bibinfo{author}{{Zonca}, E.}, \bibinfo{year}{2001}.
\newblock \bibinfo{title}{{The European Photon Imaging Camera on XMM-Newton:
  The MOS cameras : The MOS cameras}}.
\newblock \bibinfo{journal}{\aap} \bibinfo{volume}{365},
  \bibinfo{pages}{L27--L35}.
\newblock \DOIprefix\doi{10.1051/0004-6361:20000087},
  \href{http://arxiv.org/abs/astro-ph/0011498}{\tt arXiv:astro-ph/0011498}.
\bibitem[{{Vaiana} et~al.(1981){Vaiana}, {Cassinelli}, {Fabbiano}, {Giacconi},
  {Golub}, {Gorenstein}, {Haisch}, {Harnden}, {Johnson}, {Linsky}, {Maxson},
  {Mewe}, {Rosner}, {Seward}, {Topka} and {Zwaan}}]{1981ApJ...245..163V}
\bibinfo{author}{{Vaiana}, G.S.}, \bibinfo{author}{{Cassinelli}, J.P.},
  \bibinfo{author}{{Fabbiano}, G.}, \bibinfo{author}{{Giacconi}, R.},
  \bibinfo{author}{{Golub}, L.}, \bibinfo{author}{{Gorenstein}, P.},
  \bibinfo{author}{{Haisch}, B.M.}, \bibinfo{author}{{Harnden}, F.~R., J.},
  \bibinfo{author}{{Johnson}, H.M.}, \bibinfo{author}{{Linsky}, J.L.},
  \bibinfo{author}{{Maxson}, C.W.}, \bibinfo{author}{{Mewe}, R.},
  \bibinfo{author}{{Rosner}, R.}, \bibinfo{author}{{Seward}, F.},
  \bibinfo{author}{{Topka}, K.}, \bibinfo{author}{{Zwaan}, C.},
  \bibinfo{year}{1981}.
\newblock \bibinfo{title}{{Results from an extensive Einstein stellar survey.}}
\newblock \bibinfo{journal}{\apj} \bibinfo{volume}{245},
  \bibinfo{pages}{163--182}.
\newblock \DOIprefix\doi{10.1086/158797}.
\bibitem[{{Verbunt} et~al.(1997){Verbunt}, {Bunk}, {Ritter} and
  {Pfeffermann}}]{1997A&A...327..602V}
\bibinfo{author}{{Verbunt}, F.}, \bibinfo{author}{{Bunk}, W.H.},
  \bibinfo{author}{{Ritter}, H.}, \bibinfo{author}{{Pfeffermann}, E.},
  \bibinfo{year}{1997}.
\newblock \bibinfo{title}{{Cataclysmic variables in the ROSAT PSPC All Sky
  Survey.}}
\newblock \bibinfo{journal}{\aap} \bibinfo{volume}{327},
  \bibinfo{pages}{602--613}.
\bibitem[{{Warner}(2003)}]{warner2003cataclysmic}
\bibinfo{author}{{Warner}, B.}, \bibinfo{year}{2003}.
\newblock \bibinfo{title}{Cataclysmic Variable Stars}.
  volume~\bibinfo{volume}{28}.
\newblock \bibinfo{publisher}{Cambridge University Press},
  \bibinfo{address}{Cambridge}.
\newblock \DOIprefix\doi{10.1017/CBO9780511586491}.
\bibitem[{{Willingale} et~al.(2013){Willingale}, {Starling}, {Beardmore},
  {Tanvir} and {O'Brien}}]{2013MNRAS.431..394W}
\bibinfo{author}{{Willingale}, R.}, \bibinfo{author}{{Starling}, R.L.C.},
  \bibinfo{author}{{Beardmore}, A.P.}, \bibinfo{author}{{Tanvir}, N.R.},
  \bibinfo{author}{{O'Brien}, P.T.}, \bibinfo{year}{2013}.
\newblock \bibinfo{title}{{Calibration of X-ray absorption in our Galaxy}}.
\newblock \bibinfo{journal}{\mnras} \bibinfo{volume}{431},
  \bibinfo{pages}{394--404}.
\newblock \DOIprefix\doi{10.1093/mnras/stt175},
  \href{http://arxiv.org/abs/1303.0843}{\tt arXiv:1303.0843}.
\bibitem[{{Wilms} et~al.(2000){Wilms}, {Allen} and
  {McCray}}]{2000ApJ...542..914W}
\bibinfo{author}{{Wilms}, J.}, \bibinfo{author}{{Allen}, A.},
  \bibinfo{author}{{McCray}, R.}, \bibinfo{year}{2000}.
\newblock \bibinfo{title}{{On the Absorption of X-Rays in the Interstellar
  Medium}}.
\newblock \bibinfo{journal}{\apj} \bibinfo{volume}{542},
  \bibinfo{pages}{914--924}.
\newblock \DOIprefix\doi{10.1086/317016},
  \href{http://arxiv.org/abs/astro-ph/0008425}{\tt arXiv:astro-ph/0008425}.

\end{thebibliography}

%
%
%
%

\end{document}